\documentclass[a4paper,twocolumn,showpacs,prl]{revtex4}%
\usepackage{mathrsfs}
\usepackage{amsfonts}
\usepackage{amsmath}
\usepackage{amssymb}
\usepackage{graphicx}
\usepackage{float}%
\providecommand{\U}[1]{\protect \rule{.1in}{.1in}}

\begin{document}
\title{Interface-induced Topological Insulator Transition in GaAs/Ge/GaAs Quantum Wells}
\author{Dong Zhang$^{1}$, Wenkai Lou$^{1}$, Maosheng Miao$^{2}$, Shou-cheng
Zhang$^{3}$ and Kai Chang$^{1}$}
\affiliation{$^{1}$SKLSM, Institute of Semiconductors, Chinese Academy of Sciences, P.O.
Box 912, Beijing 100083, China}
\affiliation{$^{2}$Materials Research Laboratory and Materials Department, University of
California, Santa Barbara, California 93106-5050, USA}
\affiliation{$^{3}$Department of Physics, Stanford University, Stanford, CA94305}

\begin{abstract}
We demonstrate theoretically that interface engineering can drive Germanium,
one of the most commonly-used semiconductors, into topological insulating
phase. Utilizing giant electric fields generated by charge accumulation at
GaAs/Ge/GaAs opposite semiconductor interfaces and band folding, the new
design can reduce the sizable gap in Ge and induce large spin-orbit
interaction, which lead to a topological insulator transition. Our work
provides a new method on realizing TI in commonly-used semiconductors and
suggests a promising approach to integrate it in well developed semiconductor
electronic devices.

\end{abstract}

\maketitle

Time-reversal invariant topological insulators (TIs) have aroused intensive
interests in the past years, with tantalizing properties such as insulating
bulk, robust metallic edge or surface modes and exotic topological
excitations, and potential applications ranging from spintronics to quantum
computation.\cite{RMP1,RMP2,Kane,BHZ,Konig,CXLiu,Du2011,Fu,Hsieh,ZXShen,Hasan,Chadov,Lin,Franz,KYang}
Despite these successful progresses, the topological insulator materials are
still limited in the narrow gap materials containing heavy atoms, e.g.,
HgTe,\cite{BHZ,Konig} Bi$_{2}$X$_{3}$ (X=Se,
Te,...)\cite{Fu,Hsieh,ZXShen,Hasan}, transition metal oxide
heterostructure\cite{Oxide} and Heusler compounds\cite{Chadov,Heusler}. These
materials are often very different from conventional semiconductor materials
in structures and properties and are hard to be integrated in current
electronics devices that are based on well developed semiconductor fabrication technologies.

Although there are theoretical predicts about realizing TI states in
graphene,\cite{Kane, Neto,Bilayer} the main obstacle is the weak intrinsic
spin-orbit interaction (SOI) of carbon atoms. Here, instead of searching new
TI materials with exotic structures and chemical elements, we take a totally
different route: driving the commonly used semiconductors into TI states by
using the intrinsic electric field and the strains. The difficulty of this
approach lies in the fact that most of the commonly used semiconductors, such
as Si, Ge, GaAs and many others usually possess sizable band gaps and do not
have strong enough SOI. Furthermore, group IV elements such as Si and Ge have
indirect band gap, posing extra difficulty in realizing TI. Inspired by recent
theoretical works that the normal insulator\cite{Chang} can be driven into a
TI by an external electric field, our approach is to impose huge electric
field by deliberately designed heterostructures. Recently, an interesting way
of realizing topological insulating phase in a p-type GaAs quantum well by
two-dimentional superimposed potentials with hexagonal symmetry was
proposed.\cite{GaAs} Different to that work, our approach rely completely on
the material engineering at the atomic level.

\begin{figure}[tbh]
\includegraphics[width=1.0\columnwidth]{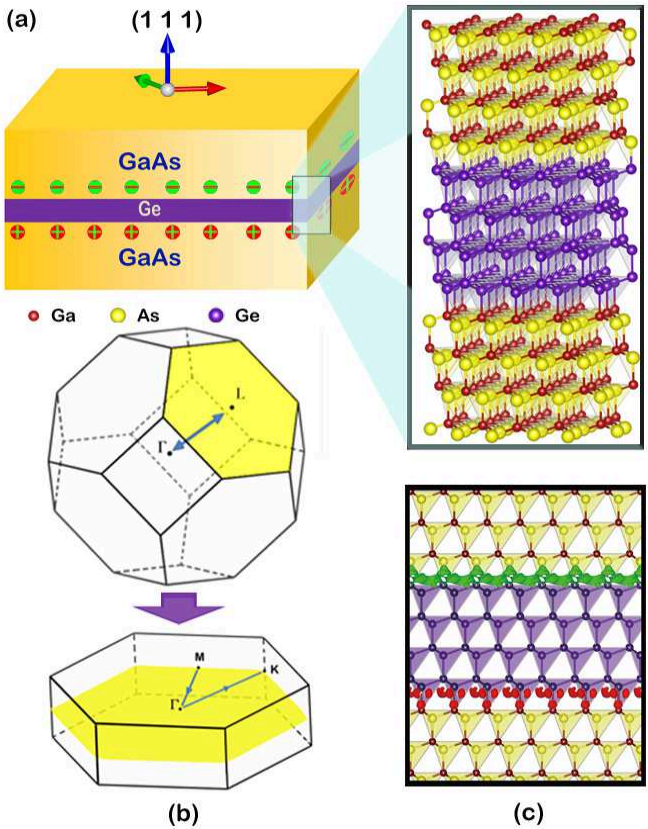}\caption{(color online)(a)
Schematic of the structure of a ultrathin Ge layer sanwiched by thick GaAs
layers (the upper-left panel). The upper-right panel amplifies the atomic
configuration of the GaAs/Ge/GaAs quantum well containing four bilayer Ge.
Notice that the Ga and As atoms locate at the opposite interfaces which leads
to a charge accumulation schematically shown in the left panel.\ (b) The
Brillouin zone (BZ) of bulk\ Ge and the folded BZ of GaAs/Ge/GaAs QW along the
[111] crystallographic direction. (c) The charge accumulation at two opposite
interfaces obtained from the first-principle calculation. The red and green
isosurfaces describe the positive and negative charge accumulations at
opposite interfaces.}%
\label{fig1}%
\end{figure}

Since commonly-used semiconductors, e.g., Si, Ge, GaAs, posses sizable bandgap
ranging from 0.8eV to 1.4eV, a huge electric field is required to closing
bandgap and even invert the conduction and valence bands. Such huge electric
field can not be generated utilizing the gate technique. However, recent
technical advances in the atomic-scale systhesis makes it possible to
fabricate high quality semiconductor and oxide heterostrutures. It provides us
abundant opportunities to create novel quantum states and emergent phenomena
at the interfaces by reconstructing charge, spin and orbital states. Very
recently, a new way to exploring topological insulating phase in
semiconductors was proposed utilizing strong piezoelectric effect at the
interface between GaN and InN.\cite{InN} The strain between these material
results in a huge polarization and electric field cross the interfaces. This
huge electric field not only can invert the conduction and valence bands, but
also generate strong Rashba SOI, eventually drive the system into topological
insulating phase. Strong strain ($\sim10\%$) in this system may cause two
opposite effects, since it drives the system into topological insulating
phase; but the release of the strain can also induce defect, vacancy and
dislocation in the samples, the density of these defects increases rapidly as
the thickness of InN layers increases, making the sample growth and
fabrication very challenging. To overcome this obstacle, it would greatly
advance the field if one can realize the topological insulator in
lattice-matched common semiconductors.

Ge and GaAs are both important materials for microelectronic and
optoelectronic device applications. Very recently, Ge/GaAs heterostructures
realized by epitaxy methods paved the way to heterostructure based devices
utilizing the band offsets, quantum size effects and band structure
modifications by electric fields. Ge/GaAs quantum structures promise the
dramatic mobility increase needed for power saving
electronics.\cite{Ge,Ge1,Ge2,Ge3} Ge/GaAs interfaces with exceedingly small
lattice mismatch posses many advantages over the strained interface. Ge layers
grown on GaAs substrate were studied because of their widespread applications
in solar cells,\cite{Ge4} metal-oxide-semiconductor field-effect
transistors,\cite{Ge5} millimeter-wave mixer diodes,\cite{Ge6} temperature
sensors,\cite{Ge7} and photodetectors.\cite{Ge8}

Considering a GaAs/Ge/GaAs quantum well (QW) grown along the polar direction
[111] (see Fig.~1), a large electric field can be induced in the sandwiched Ge
layer. In GaAs/Ge/GaAs QW, one interface consists of As-Ge bonds and the other
consists of Ga-Ge bonds. Because each As contains five electrons whereas Ga
contains only three, charge will transfer from As-Ge side to the Ga-Ge side
and a large electric field will be created and imposed on the Ge layer. The
electric field will shift the electron and the hole states to the left and
right sides of the QW and reduce their energy difference (band gap), and may
eventually invert their order.\ In addition, the electric field also induces a
considerably large Rashba SOI.

In order to demonstrate the TI transition in GaAs/Ge/GaAs, we employ two
complementary approaches, the first principles methods based on density
functional theory (DFT) and the multi-band \textbf{k$\cdot$p} theory. In the
first approach, the GaAs/Ge/GaAs structure is modeled by a series of
supercells consisting of 15 atomic bilayers in which the thickness of Ge layer
varies from 1 bilayer to 6 bilayers. (periodicity requires that the total
number of atomic layers be even). Because of the key importance of the band
gaps, we use a hybrid functional in Heyd-Scuseria-Ernzerhof (HSE) scheme, an
approach that has been proved to yield band gaps in good comparison with
experimental values for majority of semiconductor materials,\cite{PAW,HSE} as
implemented in the VASP program.\cite{VASP} Using a standard mixing parameter
of 0.25, we found the band gaps of GaAs and Ge to be 1.08 and 0.73 eV
(Ref.~\onlinecite{HSE2}). The lattice parameters $a$=$b$=$c$ are found to be
5.655 \AA ~ for GaAs and 5.657 \AA ~ for Ge. On the other hand, SOI is not
included in our DFT calculations because of the exceedingly large computing
demand while combining HSE and SOI. Instead, we adopt a 30-band
\textbf{k$\cdot$p} Hamiltonian with SOI and apply it to the GaAs/Ge/GaAs QWs.
The 30-band \textbf{k}$\cdot$\textbf{p} model was used to calculate the band
structure of commonly-used semiconductors in whole Brillouin
zone.\cite{Cardona,Richard,Kurdi}

\begin{figure}[tbh]
\includegraphics[width=1.0\columnwidth]{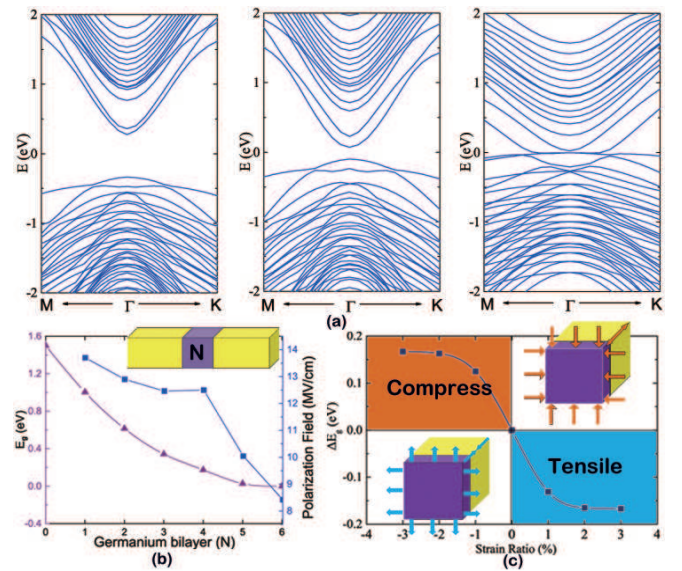}\caption{(color online) (a)
Band structures of GaAs/Ge/GaAs sandwiched structures with different Ge
portions obtained from the first-principles HSE calculations. From left to
right, two Ge bilayers, four Ge bilayers and four Ge bilayers with 3\%
in-plane tensile strain. (b) The bandgap (purple-diamond line) and the inner
polarization field strength (blue-square line) as functions of the number of
Ge bilayers. (c) The variation of bandgap $\Delta E_{g}=E_{g}^{strain}-E_{g}$
as a function of in-plane strain.}%
\label{fig2}%
\end{figure}\bigskip \ 

The proposed GaAs/Ge/GaAs structure is shown schematically in Fig.~\ref{fig1}%
(a), together with an inset showing the atomic structure of the GaAs/Ge
interfaces along the (111) growth direction. As is well known, Ge is an
indirect semiconductor with the VBM at the $\Gamma$ point and the CBM at the L
point. While growing a heterojuction along (111) direction, the symmetry are
broken and the bands are folded along the $\Gamma$-$L$ direction (see
Fig.~\ref{fig1}(b)). As a result, the CBM at the L points in bulk Ge is folded
to the $\Gamma$ point and the Ge thin layers growing in (111) direction
possesses a direct gap at $\Gamma$. The band folding is non-trivial, as we
will show below in both DFT calculation and \textbf{k$\cdot$p} model, the
breaking of the cubic symmetry also leads to a strong coupling between the
electron and the hole states, which is essential for the TI transition.

Fig.\ref{fig2}(a) presents the HSE band structures of GaAs/Ge/GaAs QW with 2
and 4 bilayers of Ge. The two-Ge-bilayer QW has a gap of 0.8 eV that is
comparable to the direct gap of Ge at $\Gamma$ point. At four Ge bilayers, the
gap decreases to 0.3 eV. The large remaining band gap is the result of strong
quantum confinement effect, which will decrease with the increasing Ge layer
thickness. In both these cases, the band structures still display normal order
in the sense that the heavy hole (HH) and light hole (LH) states are
degenerate at the $\Gamma$ point ($\Gamma_{5}$) and are lower in energy than
the electron state (E)($\Gamma_{1}$). As a matter of fact, our HSE
calculations show that the charge field itself can not drive the system into
inverted bands. As shown in Fig.\ref{fig2}(c), the band gap keep decreasing
but remain positive with up to 6 bilayers of Ge.

The driving force of the decreasing band gap is the strong electric field
imposed by the charges located at the two interfaces in GaAs/Ge/GaAs QW (see
Fig.~\ref{fig1}(c)). In principle, the strength of the electric field should
not change with increasing thickness of the Ge layer. However, because of the
finite size of our supercell, the actual model in the HSE calculations are a
GaAs/Ge superlattice, in which the GaAs region is much thicker than the Ge
region. For that reason, the electric field slightly goes down with increasing
Ge thickness. This feature is well captured by the HSE calculations as shown
in Fig.\ref{fig2}(c). While the QW contains more than 4 Ge bilayers, the HSE
calculations shows that the electric field goes down dramatically. It
decreases from 12.5 MV/cm for 4 Ge bilayer to 8.5 MV/cm for 6 Ge bilayer. The
decreasing electric field is a result of charge transfer from Ga-Ge interface
to As-Ge interface, which weakens the driving force toward band inversion.

While the bands of GaAs/Ge/GaAs QW remains in normal order with increasing Ge
thickness, we found that a slight tensile strain is enough to drive the system
into inverted bands. As shown in the upper right panel of Fig.\ref{fig2}(a),
the system exhibits an inverted band structure, in which $\Gamma_{5}$ states
are 0.1 eV higher than the $\Gamma_{1}$ state for 4 Ge bilayer with 1\%
tensile strain, and 0.2eV for 2\% strain. The actual band inversion can happen
under a much smaller strain. This slight tensile strain can be realized by
doping In atoms into GaAs host material, bending the sample, or growing the
heterostructures on a substrate with larger lattices. In Fig. 2(b), the states
around $\Gamma$ point are still denoted as E, HH and LH according to its
energy order, which is different to the notation used in previous
works.\cite{BHZ} Such an inverted band structure is a signature of the
transition to a TI state. Similar to bulk Ge, the effect of the strains on the
band gaps of GaAs/Ge/GaAs QW are quite strong. As shown in Fig. \ref{fig2}(c),
the band gap can change for about 0.2 eV with 2\% compressive and tensile
strains, providing us an effective way to control the TI transition.

Although the HSE band structure calculations show that the combination of the
strong electric field and a modest strain can invert the bands of GaAs/Ge/GaAs
QW at $\Gamma$ point, it is not sufficient to prove the TI transition. In
order to show the TI transition, both band inversion and the SOI should
present. The calculation of band structure at the hybrid functional level with
SOI for a QW system is extremely demanding on computing resources. We
therefore take another approach by constructing a multi-band \textbf{k$\cdot
$p} model Hamiltonian. The parameters of the model are carefully calibrated
with HSE calculations. Comparing with HgTe system, there are more valence
bands involved in interacting with the electron state at the $\Gamma$ point.
We find that the inclusion of 30 bands in total is sufficient to describe
complicated interband coupling at the $\Gamma$ point of the Brillouin zone of
GaAs/Ge/GaAs QWs. The most important states near the bandgap are the spin-up
and spin-down electron states ($\left \vert E,\uparrow \right \rangle $ and
$\left \vert E,\downarrow \right \rangle $), the spin-up and spin-down heavy hole
($\left \vert HH,\uparrow \right \rangle $ and $\left \vert HH,\downarrow
\right \rangle $) states. We would like to emphasize that the electron and
heavy hole subbands ($\left \vert E,\uparrow \downarrow \right \rangle $ and
$\left \vert HH,\uparrow \downarrow \right \rangle $) denote only the dominant
components of the lowest conduction and highest valence subbands and are mixed
with electron and heavy- and light-hole states due to the interband coupling
in the \textbf{k$\cdot$p} theory.

\begin{figure}[tbh]
\includegraphics[width=1.0\columnwidth]{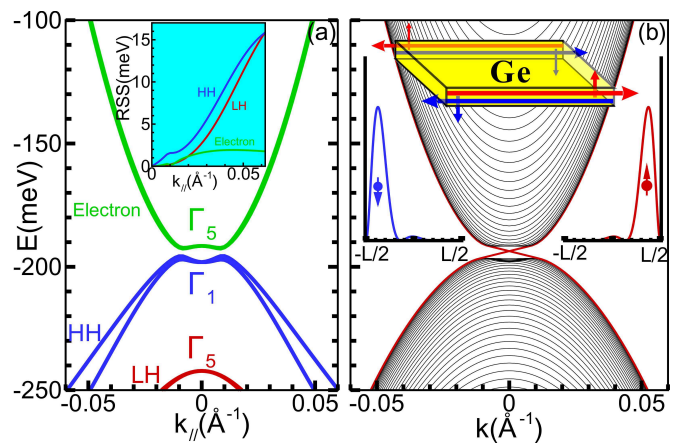}\caption{(color online) (a)
Band structures of a GaAs/Ge/GaAs QW structure with 18 \AA \ Ge layer
thickness obtained from the thirty-band \textbf{k}$\cdot$\textbf{p} model. The
inset shows the Rashba spin splitting of electron, heavy-hoel (HH) and
light-hole (LH). (b) Band structure of the quantum wire obtained by solving
the effective four-band model. The gapless edge states are showed in red line.
The central inset shows the schematic of the quantum wire and the helical edge
states. The right (red) and left peaks (blue) describe the density
distribution of the spin-up and spin-down edge states at k$_{\parallel}$=-0.01
\AA ~$^{-1}$, respectively.}%
\label{fig3}%
\end{figure}

By applying the \textbf{k$\cdot$p} theory, we first confirm the band inversion
in the GaAs/Ge/GaAs QW system. The numerical simulation shows that the
inversion happens when the thickness of Ge layer is larger than 18\AA \ and
the QW is subject to about 0.5\% tensile strain. This corresponds to about 4
Ge bilayers and is in excellent agreement with the HSE results.\ The inverted
bands are shown in Fig. \ref{fig3}(a), the electron $\Gamma_{1}$ state is
lower in energy than the valence $\Gamma_{5}$ state. Clearly, it is the result
of the fact that the highest valence subbands $\left \vert HH,\uparrow
\downarrow \right \rangle $ are heavily involved in the coupling with electronic
subbands $\left \vert E,\uparrow \downarrow \right \rangle $ near the $\Gamma$ point.

Spin-orbit interaction is essential in the transition to a TI state. The
intrinsic SOIs in both Ge and GaAs are not strong enough to creating TI
states. We found that the large interface electric field induces a
considerably large Rashba SOI splitting from the 30-band \textbf{k}$\cdot
$\textbf{p} theory for electron and hole states ($\sim$2-15meV, see the inset
of Fig. 3(a)), respectively. The magnitude is comparable with that in HgTe
QWs.\cite{Chang} This large Rashba SOI in GaAs/Ge/GaAs QW is a nature results
of the strong electric field and can be derived from the multi-band
\textbf{k}$\cdot$\textbf{p} theory. No fitting parameters are required except
the strength of the electric field, which is adopted from the HSE DFT
calculation (see Fig. 2(b)). The strong Rashba SOI in QWs provides a
controllable approach to creat TI states in GaAs/Ge/GaAs QWs. The strengths of
the Rashba SOI in the GaAs/Ge/GaAs QWs are comparable to the Rashba spin
splitting induced by an external electric field in InAs and HgTe quantum wells
\cite{Chang}. A Rashba SOI of this magnitude usually occurs only in systems
containing heavier atoms. The unusually large Rashba SOI in GaAs/Ge/GaAs QWs
is due to the strength of the polarization field, which easily exceeds 10
times the strength of an applied electric field resulting from the
state-of-art gate technique.

Next, we will demonstrate the topological insulator transition in this
GaAs/Ge/GaAs QW systems. The light-hole subbands with the dominant components
$\left \vert LH,\uparrow \downarrow \right \rangle $ are about 50$meV$ below the
electron and heavy-hole subbands ($\left \vert E,\uparrow \downarrow
\right \rangle $ and $\left \vert HH,\uparrow \downarrow \right \rangle $) (see
Figs. S2 and S3 in the online Supplymentary Materials). Since we are only
interested at the edge states inside the bulk gap ($\sim$10$meV$), the
light-hole subbands $\left \vert LH,\uparrow \downarrow \right \rangle $ are not
necessary to be explicitly included when we reduce the multi-band
\textbf{k}$\cdot$\textbf{p} model to the effective 2D \textbf{k}$\cdot
$\textbf{p} model. We downfold the 30 band model Hamiltonian to an effective
two-dimensional (2D) four-band Hamiltonian expressed in the above four basis
($\left \vert E,\uparrow \downarrow \right \rangle $ and $\left \vert
HH,\uparrow \downarrow \right \rangle $). The exact form and the derivation
process of the four-band effective 2D Hamiltonian can be found in the
supplementary Information. The contributions from the lowest and highest ten
subbands are included in this four-band reduced Hamiltonian via the L\"{o}wdin
perturbation theory \cite{Lowdin}. The explicit expression of four-band
effective two-dimenional Hamiltonian Hamiltonian in the basis $\left \vert
E_{1},\uparrow \right \rangle $, $\left \vert HH_{1},\uparrow \right \rangle $ is
\begin{equation}
H_{4\times4}^{eff}=\left[
\begin{array}
[c]{cccc}%
E_{0}+E_{1}k_{\parallel}^{2} & A_{1}k_{+} & 0 & 0\\
A_{1}^{\ast}k_{-} & H_{0}+H_{1}k_{\parallel}^{2} & 0 & 0\\
0 & 0 & E_{0}+E_{1}k_{\parallel}^{2} & -A_{1}k_{-}\\
0 & 0 & -A_{1}^{\ast}k_{+} & H_{0}+H_{1}k_{\parallel}^{2}%
\end{array}
\right]  ,
\end{equation}
where $k_{\parallel}$ denotes the in-plane momentum, and $k_{+}=k_{x}\pm
ik_{y}$, the relevant parameters $E_{0}=-0.19808$ eV, $E_{1}=0.43810\ $%
eV$\cdot%
\operatorname{\r{A}}%
^{2}$, $H_{0}=-0.19153$eV, $H_{1}=-0.20810$eV$\cdot%
\operatorname{\r{A}}%
^{2}$, $A_{1}=0.028510$eV$\cdot%
\operatorname{\r{A}}%
$.

From the above four-band effective Hamiltonian, we can start to study the
topological insulator transition in such 2D QWs. The essential feature of 2D
TI is the existence of the helical edge states near the boundary of 2D\ TI
sample. We consider a quantum wire structure with a width of 1000 \AA . The
thickness along the (111) growth direction consists of a Ge layer of 18 \AA ,
sandwiched by two 200 \AA ~GaAs layers. Fig. \ref{fig3}(b) shows the band
structure of the above quantum wire together with the density distribution of
a Kramers pair of edge states. As shown in the figure, the new energy branches
appear and sweep across the bulk gap, these states are highly localized near
the edge of the quantum wire. The spin-up and spin-down edge states with the
same in-plane momentum $k=0.01%
\operatorname{\r{A}}%
^{-1}$ along the quantum wire localize at the opposite edges, in contrast to
the chiral edge states in the integer quantum Hall effect, where the spin-up
and spin-down electron with the same in-plane momentum localize at the same
edge. The presence of these helical edge states clearly demonstrates the TI
transition in this two-dimensional GaAs/Ge/GaAs QW system.

The experimental detection the aforementioned edge states in GaAs/Ge/GaAs Hall
can be performed in the standard four terminal measurements. The edge states
can be observed at the minigap opened between the E1 and the HH1 bands.
According to our calculations using the \textbf{k$\cdot$p} model, this minigap
can be as large as 15 meV, which is already larger than the similar minigap in
InAs/GaSb QW system, another 2D TI in which the edge states has recently been
observed.\cite{Du2011} The presence of the TI state in Ge ultrathin layers can
largely advance the application of this new quantum state in existing
electronics and optoelectronics devices. It shows a considerable advantage
over other TI systems including graphene \cite{Kane}, HgTe QW,\cite{BHZ,Konig}
the Bi chalcogenides\cite{Hsieh} and the Heusler compounds
\cite{Chadov,Heusler}. GaAs/Ge/GaAs sandwiched structures are readily to be
integrated with conventional semiconductors which are already extensively used
in electronic devices \cite{Ge,Ge1,Ge2,Ge3,Ge4,Ge5,Ge6,Ge7,Ge8}. The imposed
electric field can be controlled by applying extra electric field or by
inducing holes or electrons into the QW region via a gate voltage, providing
us a direct way of manipulating the TI transition in the device. The
transition point can also be adjusted by well developed semiconductor
techniques such as alloying and doping.

\begin{acknowledgments}
This work was supported by the NSFC Grants No. 10934007 and the grant No.
2011CB922204 from the MOST of China, and M.S.M. is supported by the MRSEC
program (NSF- DMR1121053) and the ConvEne-IGERT Program (NSF-DGE 0801627).
S.C.Z. is supported by the DARPA Program No. N66001-12-1-4034. KC would like
to thank Prof. Bangfen Zhu, Qikun Xue, Zhong Fang, Xi Dai for their valuable discussions.
\end{acknowledgments}

\end{document}